\documentclass[12pt,preprint]{aastex}
\usepackage{emulateapj5}
\slugcomment{Astrophysical Journal, in press}
\shortauthors{Kronberg et. al.}
\newcommand{\bfw}{\mbox{\boldmath $w$}}
\newcommand{\bfs}{\mbox{\boldmath $s$}}
\newcommand{\bfY}{\mbox{\boldmath $Y$}}

\begin{document}
\title{A Global Probe of Cosmic Magnetic Fields to High Redshifts}

\author{P. P. Kronberg\altaffilmark{1,2}, M. L. Bernet\altaffilmark{3},
F. Miniati\altaffilmark{3}, S.J. Lilly\altaffilmark{3}, M. B.
Short\altaffilmark{1}, D. M. Higdon\altaffilmark{1}}

\altaffiltext{1}{Los Alamos National Laboratory, P.O. Box 1663, Los
Alamos NM 87545 USA; kronberg@lanl.gov, mbshort@lanl.gov,
dhigdon@lanl.gov}

\altaffiltext{2}{Department of Physics, University of
Toronto, 60 St. George, Toronto M5S 1A7, Canada;
kronberg@physics.utoronto.ca}

\altaffiltext{3}{Physics Department, Wolfgang-Pauli-Strasse 16, ETH Z\"urich, CH-8093 Z\"urich, Switzerland;
mbernet@phys.ethz.ch, fm@phys.ethz.ch, simon.lilly@phys.ethz.ch}

\begin{abstract}

{Faraday rotation (RM) probes of magnetic fields in the universe are
sensitive to cosmological and evolutionary effects as $z$ increases
beyond $\sim $1 because of the scalings of electron density and
magnetic fields, and the growth in the number of expected intersections with galaxy-scale
intervenors, $d$N/$dz$. In this new global analysis of an
unprecedented large sample of RM's of high latitude quasars extending out to 
$z\sim $3.7 we find that the distribution of RM broadens with redshift in 
the 20 $-$ 80 rad m$^{-2}$ range range, despite the (1 +$z$)$^{-2}$ wavelength dilution expected in the observed
Faraday rotation. Our results indicate that the Universe becomes increasingly
``Faraday-opaque'' to sources beyond $z \sim$ 2, that is, as $z$ increases progressively fewer
sources are found with a ``small'' RM in the observer's frame. This is in contrast to sources at $z \la$1. 
They suggest that the environments of galaxies were significantly magnetized at high redshifts,
with magnetic field strengths that were at least as strong within a few Gyr of the 
Big Bang as at the current epoch.
We separately investigate a simple unevolving toy model in which the RM is 
produced by MgII absorber systems, and find that it can approximately reproduce the observed 
trend with redshift. An additional possibility is that the intrinsic RM 
associated with the radio sources was much higher in the past,
and we show that this is not a trivial consequence of the higher radio 
luminosities of the high redshift sources.
}

\end{abstract}


\section{INTRODUCTION}

The strengths of interstellar and intergalactic magnetic fields at
earlier epochs have important implications for galaxy and structure
evolution (e.g. Mestel \& Paris 1984; Rees 1987), the propagation of
ultra-high energy cosmic rays (Sigl, Miniati, \& En\ss lin
2003, 2004; Armengaud, Sigl \& Miniati, 2005; Dolag et al. 2005), and
the feedback of magnetic energy into the intergalactic medium by
stellar winds and early massive black holes (Kronberg et al. 2001).

Faraday rotation of distant polarized radio sources is one of the few
available measurables to detect and probe extragalactic magnetic
fields. For a cosmologically distant polarized source at redshift
$z_{s}$ it is defined, in units of rad/m$^{2}$, as:

\begin{equation}
\label{eq:defRM}
RM(z_{s}) = \frac{{\Delta \chi_0 }}{{\Delta \lambda _0^{2} }} = 8.1 \cdot
10^{5} \int\limits_0^{z_s } \frac{n_e (z)B_\parallel(z)}{(1+z)^{2}}\frac{dl}{dz}dz.
\end{equation}
The RM describes the change in polarization angle $\Delta \chi_0$ with
respect to a change in wavelength squared $\Delta \lambda _0^{2}$ due to
the presence of a magnetized medium (the subscript 0 indicates
observer's frame). In Eq. (1) the free electron number density, $n_{e}$, is in
cm$^{-3}$, $B_{\|}$, in Gauss, is the line of sight component of the
magnetic field and $dl$/$dz$, in parsecs, is the comoving path
increment per unit redshift. In general the total RM of a given radio
source will be a sum of several different components: (1) a ``smooth'' Galactic
component, defined as $\rm{SRM}$, that may be assumed to  vary
with $l$ and $b$ on angular scales that are larger than the typical inter-source separation.
This also includes any metagalactic and/or local
universe $\rm{RM}$ contributions that might exist on large angular scales; (2) A
component arising from intervening discrete clouds, e.g. galaxies,
and/or a diffuse medium along the line of sight. The latter includes
filaments of cosmological Large Scale Structure (LSS). The invervening
galaxy system component should depend, statistically, only on the
intergalactic path length traversed and not on direction. The third
(3) is an ``intrinsic'' component from magnetised plasma associated
with the distant radio source and its immediate environs,
($\rm{RRM_{intr}}$), which may depend on source-intrinsic properties
and which may also evolve cosmologically; Finally (4) there are
measurement errors, which ideally should not depend on $l,b$, or $z$.

Detailed RM images of individual quasars between $z \sim 1$ and $z
\sim 2$ with companion absorption line data have established clear
examples of case (2) e.g. PKS 1229-021 (Kronberg, Perry \&
Zukowski 1992), and of case (3), e.g. 3C191 at $z_{s}$ = 1.95, which contains
intrinsic RM variations at $\sim z_{s}$ of order 2000 rad
m$^{-2}$ (Kronberg, Perry \& Zukowski 1990). These studies probed the
intervening Faraday-active gas by analysing both optical absorption
lines and Faraday rotation images. More recently, two-dimensional RM
images from larger samples of resolved high-$z$ quasar radio maps 
(Athreya et al. 1998, and Carilli  et al. 1997), also show
similarly high Faraday rotations at 2 $\la z \la$ 4, which these
authors interpreted as indicating that the intrinsic RM dominates,
i.e. case (3).

Redshift-dependence of the SRM-corrected RRM($z$) (RRM = RM - SRM)
could be due to one or both of (2) or (3) above, and in each case is
subject to the $(1 +z)^{-2}$ watering-down effect in equation (1). 
Earlier attempts were made to detect a $z$-dependence of quasar RM's 
(Rees \& Reinhardt 1972, Kronberg \&
Simard-Normandin 1976) using RM data on samples of
extragalactic radio sources. All of these showed some evidence for an
increase in the observed RM of quasars at $z \gtrsim $ 1. With
somewhat better RM data, and with optical absorption line data for
some quasars, it was found that high column density optical and HI
absorption at intervening redshifts correlated with higher levels of
observed RM (e.g. Kronberg \& Perry 1982, Kronberg, Perry \&
Zukowski 1992, Oren \& Wolfe 1995). This allowed some first
estimates of magnetic field strengths in distant intervening galaxy
systems.

Welter et al.(1984) found a clear growth of the overall width (variance)
of $|$RRM$|$($z$) in a 116-quasar RM sample, recently confirmed 
in a smaller sample by You et al. (2003). Welter et al. also developed  
mathematical frameworks for connecting RM correlations to components (2) and
(3). They tentatively favored (2) over (3) at redshifts up to $z \sim$
2. A framework has also been developed by Kolatt (1998) for an
RM-based probe of the primordial magnetic field spectrum, related to
cases (2) and (3). More recently, optical galaxy and RM data were
combined to undertake the first magnetic field probe of LSS filaments
in the local universe (Xu, et al. 2006). Magnetic fields in the
cosmic voids of LSS have not yet been detected.

This paper analyses an optimized subset of a new, much expanded sample of
900 extragalactic RRM's with measured redshifts up to $z$ $\sim$
3.7. These also have improved determinations of the (smoothed) Galactic and
local universe foreground SRM contribution. We focus on the overall {\it
distribution} of RRM's as a function of redshift as a diagnostic of
magnetic fields in high redshift systems, in particular the little-explored 
$z$ - dependence of $|$RRM$|$ at $|$RRM$| \la$ 100 rad m$^{-2}$. This is distinct 
from previous investigations, summarized above, of the overall width, or variance
of the RRM distributions.  

The rest of the paper is organized as follows.
In Section \ref{data:sec} the 
RM dataset is briefly described and in Section \ref{bias:sec} we
discuss the selection and optimization of the RM sample, 
as well as tests for unwanted selection effects.
The key evidence for a redshift dependence of the RRM distribution
is presented in Section \ref{rrmz:sec}. We interpret the observed effects in 
Section \ref{models:sec}.
In general terms, an increase in the width of the RM distribution requires the presence of 
substantial amounts of magnetized plasma at high redshift. As one hypothesis, we construct 
a model in which the RM arises due to
the presence of magnetized clouds traced by MgII absorption systems 
(\ref{subsec:MC}). In section 6 we discuss the complexities of detecting an all-pervading, 
intergalactic magnetic field in the presence of other extragalactic RM contributions.  
Our conclusions 
are summarized in Section 7. Throughout the paper we assume a concordance 
cosmology with H$_{0}$ = 70 km s$^{-1}$Mpc$^{-1}$, $\Omega_{m}$ = 0.25, 
$\Omega_{\Lambda}$ = 0.75.

\section{The Dataset} \label{data:sec}

The expanded RM data include new linear polarization measurements
by one of us (PPK), which favored quasars at larger
redshifts, plus additional data from the literature. Our new sample of
901 quasars and radio galaxies is sufficiently large that we are able
to choose an optimized subset (by Galactic $l$,$b$ location) of 268
objects distributed up to z $\sim$ 3.7. This subset exceeds the previous
all-sky 116 quasar sample of Welter et al. (1984) by a factor of
2.3. In addition, the average quality of RM measurement is improved, as is also
the Galactic foreground correction, which used a new expanded sample of 1566 
extragalactic radio source RM's, mostly at $|b| >$ 5$^{\circ}$.

Accurate subtraction of the foreground Galactic SRM is important
for obtaining the purest possible isolation of the extragalactic RRM.
The SRM was estimated using a Bayesian, Gaussian process
formulation applied to the larger all-sky sample of 1566 extra-
galactic radio source RMs (Short, Higdon and Kronberg, 2007a,b) which
gives an SRM estimate for any $(l,b)$ location. 

\begin{figure*} 
\plotone{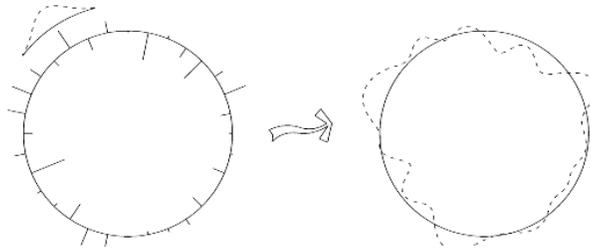}

\caption{ A realization of a process convolution model on the unit  
circle.  The left panel shows
   knot locations and values (positive and negative),
   along with a smoothing kernel.  The right panel shows the resulting
   Gaussian process realization
   obtained by a convolution of the knot values with the smoothing  
kernel. 
\label{fig:schematicfig}}
\end{figure*}

In that paper, a fairly general Gaussian process (GP) model for the  
surface of the unit
sphere  ${\cal S}$ with great circle
distance metric $d(\bfs_1,\bfs_2)$ is derived by taking a collection  
of uniform, regularly
spaced knot locations $\bfw_1,\ldots,\bfw_J$, and assigning to each  
of these
locations a knot value $x_1,\ldots,x_J$ which are assumed to have iid  
$N(0,[J \lambda_x]^{-1})$
distributions, where iid denotes ``independent and identically  
distributed.''
Convolving these knot values with a simple smoothing kernel $k(\cdot) 
$ then
results in the GP model
\begin{equation}
\label{eq:z}
    u(\bfs) = \sum_{j=1}^J x_j k(d(\bfs,\bfw_j)), \hspace{0.15in}  
\bfs \in {\cal S}.
\end{equation}
Figure \ref{fig:schematicfig} shows an example where ${\cal S}$ is  
the unit circle.
As $J \rightarrow \infty$, the process $u(\bfs)$ quickly converges  
to a stationary
GP.

The convolution kernel $k(\cdot)$ is taken to be
a normal density whose width is to be estimated.
A recursive tessellation algorithm is used to
distribute $J=2562$ knots over the unit sphere, giving a neighboring
knot-to-knot distance of approximately $2\pi/80$.
The spatial SRM field $u(\bfs)$ is given by (\ref{eq:z}) which
requires that the knot values $x_j$ and the kernel width be estimated
from the radio source RMs.
The $N=1566$ observations taken at locations $\bfs_1,\ldots,\bfs_N$
are modeled as
\[
    \bfY(\bfs_i) = u(\bfs_i) + \epsilon_i,\,i=1,\ldots,N
\]
where the errors are independent with $N(0,[\omega_i \lambda_\epsilon] 
^{-1})$ distributions.
The $\omega_i$'s, which modify the error precision, account for
the possibility that certain observations have
been altered by individual source RM anomalies, interveners, very  
small scale local
effects of the Milky Way, etc.  When $\omega_i < 1$, the observation  
has been altered
and is downweighted for the purpose of estimating the SRM field;
when $\omega_i = 1$, the observation is assumed to be free of any of  
these altering effects.

From the analysis of Short et al.~(2007a), the proportion of the  
observations identified as altered ($\omega_i < 1$) is 23\%.  The resulting  
posterior distribution from this model formulation, which accounts for (a) uncertainty  
regarding the spatial dependence in the SRM field, (b) the classification of observations  
as altered or not, and (c) observation noise, was sampled using Markov chain Monte  
Carlo.  The model formulation is described in detail by Short, Higdon \& Kronberg  
(2007a). Compared with previous methods that averaged the neighboring unedited
RM($l$,$b$), and then iteratively remove ``outliers'' (e.g. Simard-Normandin \& Kronberg 1980 and other 
papers since then), the more 
statistically formal method used here is free of ad hoc criteria, for example on the decision of
when to delete an outlier. Another important advance is that our model is with the
data to produce the best estimate of the kernel 
width, rather than having it estimated, or guessed a priori.       

Because this paper focuses on links between source redshift and RRM, it is 
important to search for possible bias effects of 
Galactic sky location on the $z$-distribution of the sources. As expected, the
Galactic ($l$,$b$) locations of the RRM's do not correlate with source
redshift. Similarly, neither the amount of foreground RM correction, SRM, 
nor the uncertainty in the SRM show any trend with source redshift, after we 
isolated the optimum ($l$,$b$) zones as described below.  

\begin{figure*} 
\plotone{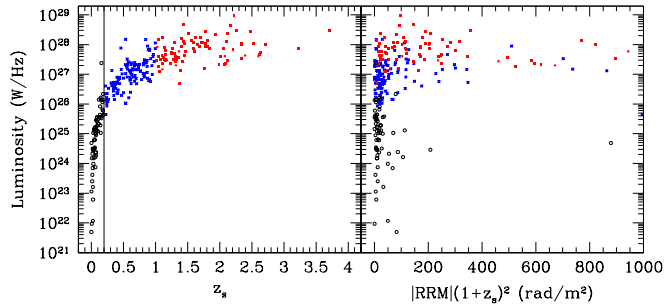}
\caption{Left
panel: Radio power at 2.7 GHz against redshift for the sample used in
our analysis. The solid line marks the subsample ($z<0.2$) used to
test for a correlation between luminosity and intrinsic RRM. Right
panel: Luminosity against RRM values transformed to the radio source
rest frame (multiplied with $(1+z_{s})^{2}$). The empty circles
are for the subsample at $z < 0.2$ that was used to test for a correlation between
luminosity and RRM, the stars are those with 0.2 $\leq z < 1.0$
and the solid squares are in the range 1.0$ \leq z < 3.7$.
\label{fig:plotLum}}
\end{figure*}

\section{Optimization of the dataset and inspection of bias effects}
\label{bias:sec}

To ensure that other biases have not been introduced in the data selection we
address three important issues that could potentially affect our
statistical analysis. First, although the subtraction of the Galactic foreground is
optimized, the quality of the SRM subtraction removal is ($l$,$b$)
dependent, being generally less accurate at the lower Galactic
latitudes. This leads to Galactic zone$-$based admission criteria for
our sample, elaborated in section \ref{sec:lbopt}.

Secondly, our sample of radio sources is, typically, flux limited so
that higher redshift objects tend to have have higher luminosities. Thus we also
investigate the RRM$ - L_{\rm radio}$ correlation to test for possible
luminosity selection effects that could prejudice our search for a RRM
- redshift relation.

Thirdly, errors in the individual RM measurements (separate from the Galactic foreground
correction uncertainty) could conceivably
exhibit a systematic redshift dependence, e.g. if the polarized
signals were much weaker at larger redshifts. On investigation we find no significant 
correlation with redshift of the uncertainty of the individual RM determinations.

\subsection{Optimal isolation of true extraglactic \rm{RM}'s}

\label{sec:lbopt} To further minimize the uncertainties due
to the subtraction of the Galactic component, 
$\rm{SRM}$, we examined how the form of the overall
$\rm{RRM}$ distribution varies as a function of the ($l$,$b$) region
on the sky.  We find that the width of the distribution
$\mathrm{N}$($\rm{RRM}$) for the smallest RRM's decreases steadily as we 
raise the lower limiting latitude from $|b|$=10$^{\circ}$(lower boundary) up to
$\left|b\right|\approx$60$^{\circ}$, but thereafter asymptotes to a
stable value. In two dimensions, we find that the boundary can be
lowered at some longitude zones without increasing the Galactic
dispersion. An optimal subsample of 268 objects for which the
influence of SRM is minimized was thus obtained by accepting all
sources at $\left|b\right|\geq $60$^{\circ}$ as above, plus those
having $b > 45^{\circ}$ and $l=150-360^{\circ}$, and those with $ b
<-50^{\circ}$ at $l=120-180^{\circ}$.

\subsection{Checks for luminosity and other unwanted
systematic effects on \rm{RRM}($z$)}

Given that the radio luminosities of the sample span about six orders of
magnitude, we can test whether the changing mean radio
luminosity with redshift is likely to be significant.  For the 268
radio source sub-sample we compared the rest-frame
(``$k$-corrected'') 2.7 GHz radio luminosity, $L_{2.7}$, with the {\it
maximum} rest-frame $\rm{RRM_{intr}}$. The latter assumes that {\it
all} the $\rm{RRM}$ comes from magnetized plasma in the vicinity of
the source, that is we multiply the observed $\rm{RRM}$ by $(1+z_s)^2$
(eq. [1]).

At $z \la 0.2$, where neither cosmological evolution nor intersection
of the line of sight with an intervenor are likely, the radio
luminosities cover more than five orders of magnitude
(Figure \ref{fig:plotLum}). However, we find no evidence for any
correlation between $|\rm{RRM_{intr}}|$ and $L_{2.7}$ (see
Figure \ref{fig:plotLum}).  To quantify this statement we
parameterize the distribution in $\rm{RRM_{intr}}$ using a Lorentzian
function

\begin{equation}
\label{eq:Lorentzian}
     f(\rm{RRM};\Gamma/2)=\frac{1}{\pi}\frac{\Gamma/2}{(\Gamma/2)^{2}+RRM^{2}},
\end{equation}
 where $\Gamma/2$ is the width at half maximum, and we test for a
$\Gamma/2$ - $L_{2.7}$ relation with

\begin{equation}
\Gamma/2=w_{0}(\mathrm{L_{2.7}}/10^{21}\mathrm{W Hz^{-1}})^{\gamma}.
\end{equation}

The best fit parameters for this relation at $z <$ 0.2 are:
$w_{0}=36_{-24}^{+57}\mathrm{rad/m^{2}}$ and
$\gamma=-0.11_{-0.10}^{+0.12}$, confirming that there is no effect of $L_{2.7}$ on
$\left|\rm{RRM_{intr}}\right|$. The right hand panel of
Figure \ref{fig:plotLum} shows this graphically. Sources with lines
of sight used to test for a correlation are represented as black
dots. The blue stars and red squares are for lines of sight from sources at
higher redshift. Due to the $(1+z_s)^{2}$ transformation they produce
much larger $\left|\rm{RRM}\right|$ values.

In conclusion, our test confirms the lack of any obvious $L_{2.7}$ - RRM
relation. Note, however, that we cannot exclude the possibility that the highest
redshift and most luminous sources could be affected by strong
evolutionary effects due to the presence of enhanced magnetic
fields when compared to their low redshift/low luminosity counterparts.

\section{Analysis of the RRM-redshift behavior} \label{rrmz:sec}

Because of the statistical nature of the RM measurements and the
present limited availability of supplementary optical data on individual
lines of sight, the most powerful diagnostic that we have is the {\it
form} of the observed distribution $\mathrm{N}($RRM$,z)$, as a
function of redshift.

To illustrate for the case of a source-intrinsic RM, if this were
$z$-independent the observed $\rm{RRM}$ should monotonically decrease
with increasing source redshift, due to the strong $(1+z)^{-2}$ term
(eq.($\ref{eq:defRM})$). This means that any variations in N($\rm{RRM}$, $z$) at
large $z$ will arise from a competition between genuine evolutionary
effects and the $(1+z)^{-2}$ reduction. Thus, an interplay of
different effects at $z \ga$ 1 might be expected. 

On the other hand, in the case of
intervening systems at low redshift their contributions to
$|\rm{RRM}|$ will be statistically invisible until there is a significant probability of
intersecting an intervenor, at which point they will rise. Then at some
sufficiently large $z$, the effect of incremental RM intersection
depth may be overwhelmed by the $(1+z)^{-2}$ term, which would
tend to flatten any increase in the observed RRM. It may subsequently
increase again if a sufficiently strong evolutionary increase sets in again
at still larger redshifts. These possibilities set the context for our
analyses in Section \ref{models:sec}. 

In the following we use a non-parametric approach,
counting fractions of lines of sight $f_{i}$ below a varying threshold
$\rm{RRM_{i}}$ as a function of $z$ (see Welter et. al 1984). We then 
apply a Kolmogorov-Smirnov test to examine the statistical
significance of the redshift dependence in our data. 

\subsection{Evidence for Redshift Evolution}

\begin{figure*}
\plotone{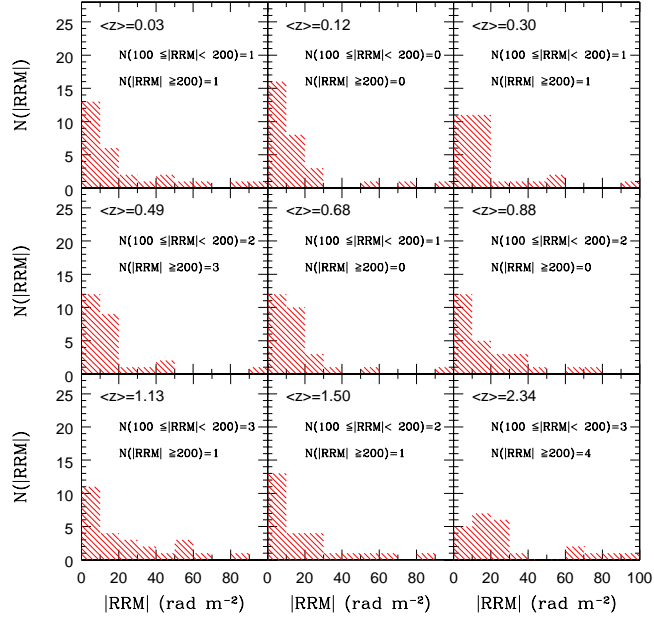}
\caption{$\left|\rm{RRM}\right|$ histograms of 268 lines of sight for
different redshift bins having approximately equal numbers of lines of
sight per bin. Redshift increases from the upper left to the lower right
panel. The distribution in the highest redshift bin ($z \ga$ 1.8) is
characterized by a significantly decreased ``peak'' near $|$RRM$|$ =
0, and a corresponding significant broadening of the ``low-$|$RRM$|$ '' peak 
to higher $|$RRM's$|$. 
\label{fig:RRMhisto} }  
\end{figure*}

Figure~\ref{fig:RRMhisto} shows the observed distributions of
$\rm{RRM}$ for 9 different redshift bins, each containing about 29
sources. At redshifts below $z \sim 1$ the distributions are
characterized by a sharply defined mode at RRM\,$\approx$\,0 plus a
broader $\rm{RRM}$ component. Beyond this redshift, and especially at
$z > 2$, the low-RRM component tends to get redistributed to larger
$\rm{RRM}$ values.

\label{subsec:analz} 

\begin{figure*}
\plotone{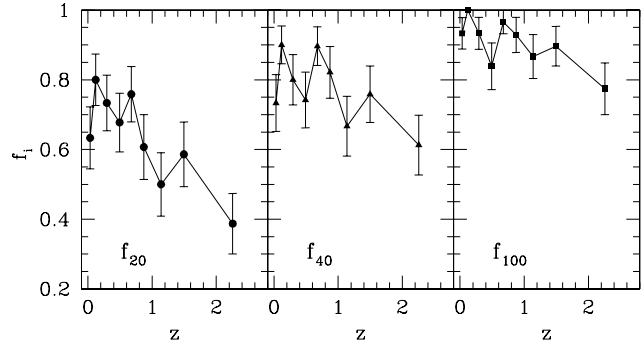}
\caption{Fraction of lines of sight $\rm{f_{i}}$
below a threshold $\rm{RRM_{i}}$ as a function of redshift. Circles,
triangles and squares show $\rm{f_{20}}$, $\rm{f_{40}}$ and
$\rm{f_{100}}$ respectively. Errors are calculated by randomly drawing
lines of sight and calculating the r.m.s. of the resulting $f_{i}$
distributions.\label{fig:nonpar}}  
\end{figure*}

This effect is shown quantitatively in Figure $\ref{fig:nonpar}$,
where the fractions of lines of sight $f_{i}$ below a varying
threshold $\rm{RRM_{i}}$ are plotted as a function of redshift. The
quantity $f_{20}$, representing the fraction of $|$RRM$|$'s below 20
rad m$^{-2}$, decreases significantly with redshift, from 72 $\%$ in
the lowest two redshift bins ($\bar{z}=0.08$) to 39$\%$ in the highest
redshift bin ($\bar{z}=2.34$). Similar, but progressively weaker
trends occur for $f_{40}$ and $f_{100}$ respectively. The highest
redshift bin shows the smallest value of $f_{i}$ for all three
thresholds. These trends indicate a clear evolutionary pattern which is most
apparent at $|$RRM$| \la 40$ rad m$^{-2}$ in the {\it observer's} frame.

\begin{figure*}
\plotone{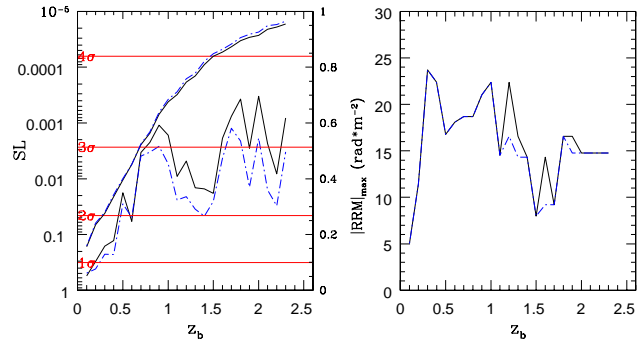}

\caption{Left panel: Significance level (SL) of the KS test that the
distributions of the RRM's below and above $z_{b}$ are not drawn from
the same parent distribution. The black solid line is with no
$\left|\rm{RRM} \right|$ cut and the blue line is for lines of sight
with $\left|\rm{RRM} \right| < 200$ $\rm{rad/m^{2}}$. The dashed
dotted lines give the fraction of lines of sight which are below
$z_{b}$ (right-hand axis). Right panel: The RRM value at which the
normalized cumulative distributions, N(RRM) from the KS-test most differ, as a
function of $z_{b}$. 
\label{fig:KStest}}
\end{figure*}

We confirm the reality of this redshift dependence with a
Kolmogorov-Smirnoff test. The entire sample was divided into two,
below and above $\rm{z_{b}}$ and we compare the two normalized cumulative
distributions of the absolute value of the $\rm{RRM}$, N$(|\rm{RRM}|)$
in Figure \ref{fig:KStest} (left panel).
For $\rm{z_b}\sim 1.8$, the samples are different at the 99\% significance
level. The $\rm{RRM}$ at which the maximum perturbation in the
N$(|\rm{RRM}|)$ occurs is typically between 10-25 rad m$^{-2}$ (Figure \ref{fig:KStest},
right panel). This shows that the clearest evolutionary signal comes from 
a broadening of the low-$|$RRM$|$ peak, rather than from
outliers in the high RRM tails of the distribution.

The ``migration'' of some sources 
near $|\rm{RRM}|$ = 0 to wider wings works {\it oppositely} to the expected
(1 + $z$)$^{-2}$ decrease. It clearly demonstrates that there is a
global evolutionary effect in Faraday rotation. More precisely,
as $z$ increases up to and beyond $\sim$ 2, there is a gradual broadening 
of the distribution of RM's and a corresponding ``depopulation'' of the lowest $|$RRM$|$
bins, especially those $\leq$ $|$20$|$ rad m$^{-2}$. 
In effect, at higher redshifts the Universe becomes progressively
``Faraday opaque''  as fewer sightlines
are able to ``escape'' an enhanced Faraday
rotation at the higher $z$'s.
The extra Faraday rotation is produced at $z > 1$ and is consequently greater in 
the Faraday rotating rest frame than what we observe, by a factor of $(1+z)^2$. 
For example, $|$RRM$|$'s of 20 to 100 in Fig. $\ref{fig:RRMhisto}$ 
become 80 to 400 rad m$^{-2}$ at $z = 1$ and 360 to 1200 rad m$^{-2}$ if the Faraday
rotation originates at $z$ = 2.5.

\begin{figure*}
\plotone{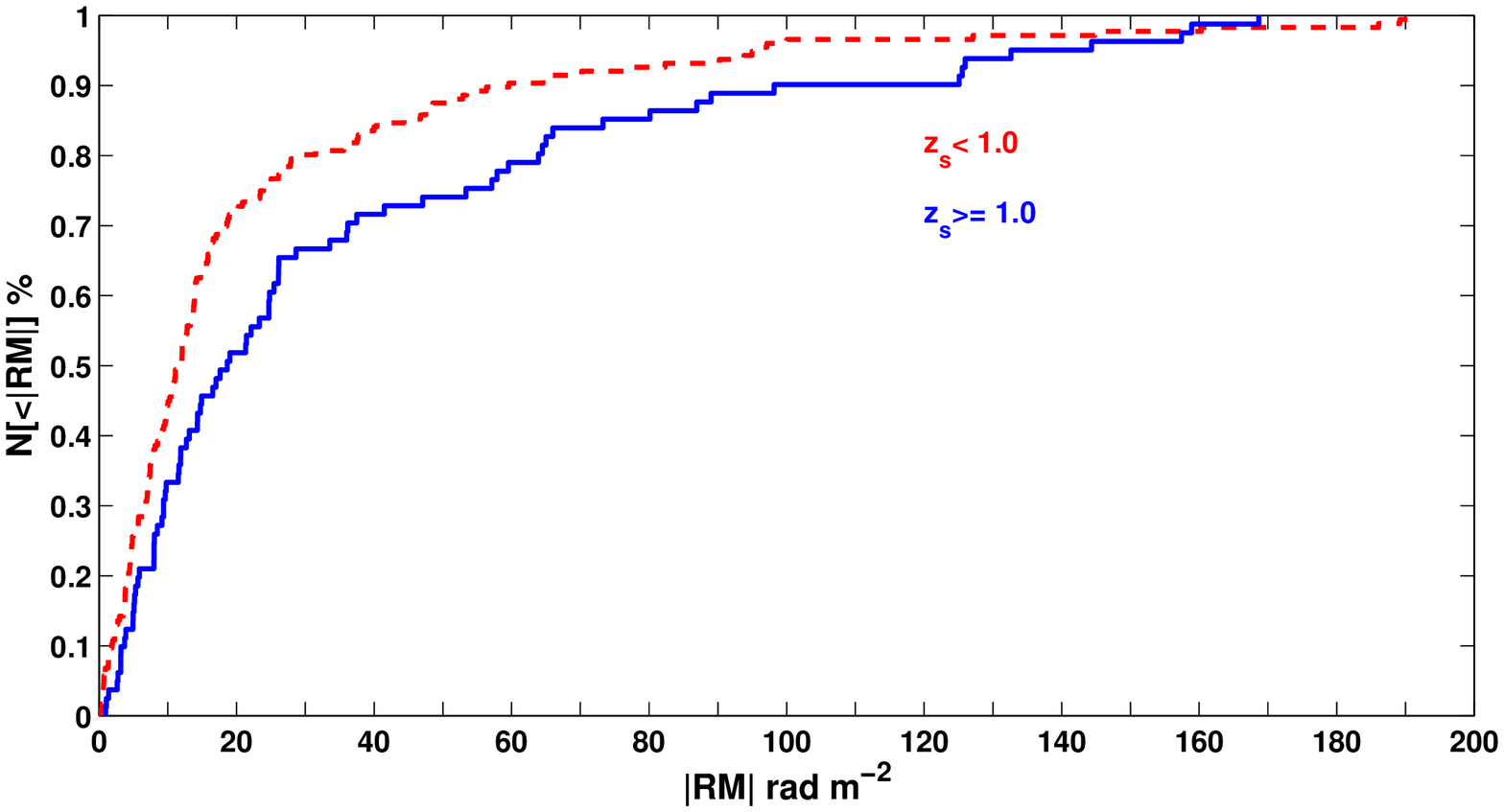}

\caption{A comparison of the normalized cumulative counts of N$(< |\rm{RRM}|)$ vs. $|\rm{RRM}|$ shown separately 
for sources having z $<$ 1 (upper curve) and z $>$ 1 (lower curve). The clear separation between these
two curves shows that excess RRM's introduced in the higher redshift subset are typically in the range 
$\left|\rm{RRM} \right|$ $\sim $ 20 to $\sim $ 80 rad m$^{-2}$ in the observer's frame. 
\label{fig:cumpdf}}
\end{figure*}

Another striking illustration of this redshift effect is shown in Figure $\ref{fig:cumpdf}$, 
which compares the normalized cumulative counts N($ < (|\rm{RRM}|)$ vs. $|\rm{RRM}|$ above and below 
z$_{b}$ = 1. Evolution in the high- $|\rm{RRM}|$ 
tails remains to be better specified in future, larger samples at the highest redshifts, and we do not attempt to
quantify it here. What is clear is that a significant evolution in the observed $|\rm{RRM}|$ at modest RRM's begins
to set in beyond $z$ $\sim $ 1.0.

\begin{table*}
    \centering
    \caption{Parameter values for different RRM models.}
        \begin{tabular}{l|l|l|l|l}
        \hline
        \hline
        & & & \\
        model & $\sigma_{noise}$ ($\rm{rad/m^{2}}$) &  $\Gamma_{intr}/2$($\rm{rad/m^{2}}$) & $\sigma_{cloud}$ ($\rm{rad/m^{2}}$)  \\
        & & &   \\
        \hline
        no intervening systems & $13_{-3}^{+4}$ & $21_{-6}^{+7}$ & -  \\
        \hline
        MgII systems with $W_{r} > 0.02 \rm {\AA}$ & $8_{-2}^{+4}$  & $7_{-4}^{+6}$ & $60_{-15}^{+20}$ \\
        \hline
        MgII systems with $W_{r} > 0.3 \rm {\AA}$ & $9_{-2}^{+4}$  & $7_{-4}^{+6}$  & $115_{-30}^{+45} $  \\
        \hline
        \hline

        \end{tabular}
    \label{tab:chival}
\end{table*}

\section{Quantitative modelling to estimate magnetic strengths 
in high redshift systems} \label{models:sec}

In the following two subsections we take different approaches to the analysis
and interpretation of the data. Each draws from the analysis
framework developed in Welter et al. (1984).

First, in Section \ref{subsec:MC}, we relate these results to recent quasar optical
absorption line data, given earlier evidence for an RM-absorption line 
association (e.g. Kronberg \& Perry 1982, Welter et al. 1984, Oren \& Wolfe 1995) 
and the {\it a priori} expectation that high column density absorption line systems will
have an effect on RRM. We make model predictions of the N(RRM, $z$) behavior
based on previous MgII absorption line studies of
quasars up to $z \sim $2.3. In order to keep the number of free model
parameters appropriate to the current state of the RRM data we assume
no local cosmological evolution in the RRM intervenor systems, and we
also exclude the very high RRM outliers from the analysis. We will
show that, even in the absence of $z$-evolution and other
contributions, MgII intervenor systems can have a recognizable
influence on the observed RRM behavior described in \S 4.

In Section \ref{sec:decline}, we interpret the broadening in the distribution of N$($RRM$)$ at
$z \ga$ 1.5 to draw general conclusions about the strength of early
universe magnetic fields in galaxy systems up to $z \sim 3.5$.

\subsection{RRM Intervenor model based on MgII absorber statistics} \label{subsec:MC}

In this section we explore the possibility that MgII absorption
systems are magnetized: Given their redshift distribution from QSO
absorption line studies we investigate whether they can explain the
observed statistical properties of the RM data and if so, what the
implications are for their magnetic properties. Due to the poor
statistics in the high-RM-tail of the distribution here we consider
only lines of sight with an observed RRM value smaller than
$\left|200\right|$ rad m$^{-2}$.  Following Welter et al.  (1984) we
calculate the probability distribution function $P$(RRM,$z_{s}$) for
an observed RRM value of a source located at redshift $z_{s}$, as 

\begin{equation} 
\label{fprob:eq} P(\rm{RRM},z_{s})= \sum_{n=0}^{n_{max}}
q_{n}(z_{s})P_{n}(\rm{RRM},z_{s}).  \end{equation} 
 
Here $P_{n}$ is the
normalized probability distribution function of RRM for a line of
sight to a source at redshift $z_{s}$ passing through $n$
intervenors. In addition, $q_{n}(z_{s})$ is the probability of having
$n$ such intervenors along the line of sight which is given by
Poisson's statistics: 
 
\begin{equation} \label{poisson:eq}
q_{n}(z_{s}) = (n!)^{-1}\nu_{s}^{n}e^{-v_{s}}, 
\end{equation} 

with 

\begin{equation} \label{nu:eq} 
\nu_{s}=\int_{0}^{z_s}\frac{dN}{dz}\; dz
\end{equation} 
 
the mean number of intervening system out to $z_{s}$
calculated using the MgII absorber distribution, $dN/dz$, described
below.  In addition to the effects of intervenors we also allow for
contributions to the observed RRM from an intrinsic component and
measurement errors, the latter dominated by uncertainties in the
removal of the Galactic contribution.  As a result,
$P_{n}$(RRM,$z_{s}$) is given by the convolution of the probability
distribution functions associated with each individual component,
namely 
\begin{equation} \label{eq:pn} 
P_{n}(\rm{RRM},z_{s})=
P_{\it noise}\ast P_{\it n,interv}(z_{s}) \ast P_{\it intr}(z_{s}), 
\end{equation}
where for both the intervenor and intrinsic component we have
explicitly indicated the redshift dependence.

In order to model $P_{n,interv}(\rm{RRM},z_{s})$ we assume that each
intervenor can be described as a cloud characterized by a number
density of free electrons $n_{e}$, a size $L$ and a randomly oriented
magnetic field $B$ with a coherence length $l_{C}$. Then for each
intervenor we can define a probability distribution function,
$P_{cloud}(\rm{RRM},z)$, given by a Gaussian distribution with $\sigma
(z)=\sigma_{cloud}(1+z)^{-2}$, where $\sigma_{cloud} \propto
n_{e}\,B\,l_{C}\,(L/l_C)^{1/2}$, i.e. has units of RM. The contribution from $n$
intervening systems is then given by the expression 

\begin{eqnarray} \label{pinterv:eq}
P_{n,{\mathop{\rm int}} erv} (RRM,z_s ) = \nonumber \\ 
AC_{n} \left[ {\int\limits_0^{z_s } {P_{cloud} (RRM,z)\frac{{dN}}{{dz}}(z)} } \right], 
\end{eqnarray}

\noindent 
which is the convolution (${\cal C}_{n}[\cdot]$) of
$n$ identical clouds distributed up to redshift $z_s$ according to
$dN/dz$, and $A$ is a normalization factor. Note that expression
\ref{pinterv:eq} is valid even if $l_C\sim L$, provided that the
magnetic field is randomly oriented from cloud to cloud and that the
analysis is applied to a large number of RRM's associated with
different lines of sight.

As for the intrinsic component, typically characterized by a
non-Gaussian tail, we find it appropriate to use a Lorentzian distribution
(eq. $\ref{eq:Lorentzian}$) which takes better account of the outlying
RM values. Assuming for simplicity no source
evolution, the latter can be fully characterized by a constant rest
frame half width at half maximum, $\Gamma_{intr}/2$, which translates
into an observed $\Gamma(z)=\Gamma_{intr}\,(1+z)^{-2}$.
Finally, $P_{noise}$, the combination of the observational error in RM and 
the uncertainty in the SRM can be represented by a
Gaussian width, $\sigma_{noise}$, which is independent of redshift.

In the following we carry out two separate analyses, one restricted to
strong absorbers, i.e. those with an equivalent width $W_{\rm r} \ge
0.3\,\rm{\AA}$, and a second including weak absorbers, i.e. those
with an equivalent width $0.02\, \rm{\AA} \leq W_{\rm r}\leq 0.3\,\rm
{\AA}$. 
For simplicity all the absorbers are characterized by the
same $\sigma_{cloud}$, independent of redshift and equivalent width
(or underlying column density).

For the weak MgII absorbers we use the function $dN/dz$ obtained by
Churchill et al. (1999) who investigated HIRES/Keck spectra of 26 QSOs
in the redshift range 0.4 $\leq$ z $\leq$ 1.4 .  They obtained the result
\begin{equation}
\frac{dN_{weak}}{dz} = (0.8\pm0.4) \,(1+z)^{1.3\pm0.9}.
\end{equation}
Similarly, for the strong MgII absoption systems we utilize
the results of Nestor et al. (2005) who, for the range
$0.4 \leq z \leq 2.3 $, find
\begin{eqnarray} \nonumber 
&&\frac{dN_{strong}}{dz} = 1.001 \, (1+z)^{0.226} \\
&\times& 
\left[\exp\left(\frac{-0.3}{\alpha\,(1+z)^\beta}\right)
- \exp\left(\frac{-6.0}{\alpha\,(1+z)^\beta}\right)\right],
\end{eqnarray} 
where $\alpha=0.443$ and $\beta=0.634$.
Note that the different distributions of weak and strong intervenors,
which enter the Poisson statistics (\ref{poisson:eq}) through the
parameter $\nu_s$ defined in eq. (\ref{nu:eq}), imply different
values for the redshift at which an absorber is first encountered.

Using the probability distribution function in eq. \ref{fprob:eq},
with the specific functional form detailed above, we now present
the results of a maximum likelihood analysis which yields the parameters that best describe the data.

We begin by considering the influence of an RRM component due to 
strong MgII absorbers, in
addition to an intrinsic component and Galactic foreground correction errors.  The value
of the parameters that produce the best fit to the data, in units rad
m$^{-2}$, are

$\sigma_{cloud}=115_{-30}^{+45}$,

$\Gamma_{intr}/2=7_{-3}^{+6}$

and 

$\sigma_{noise}=9_{-2}^{+4}$. \\
We note that, intriguingly, the typical derived value for $\sigma_{cloud}$ is not
very different from an RM observed for a typical 
line of sight through a (low $z$) spiral galaxy.

\begin{figure*} 
\plotone{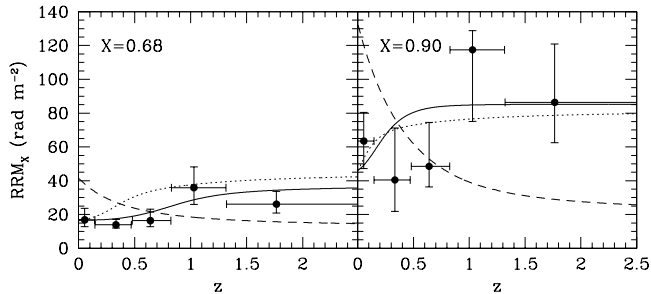}
\caption{Quantiles RRM$_{0.68}$ (left panel) and RRM$_{0.90}$(right panel) as a function 
of redshift are shown for the observed RRM data and different models. The solid lines show 
a model with strong (W$_{\rm r} > 0.3 \rm{\AA}$ ) MgII absorbers. The dotted
lines shows a intervenor model where the weak ( $0.02\rm{\AA} < W_{\rm r} \leq 0.3\rm{\AA}$) 
absorbers are included for comparison. The dashed lines show a model incorporating 
SRM removal error and a non-evolving intrinsic contribution.  Each
bin contains about 51 lines of sight, and we show the median
redshifts. Errors in the quantiles are calculated with the bootstrap
method, with the 1$\sigma$ confidence interval shown. The derived
parameter values are summarized in Table 1.\label{fig:quantiles}}
\end{figure*}

In order to illustrate how our model for $P(\rm {RRM},z_{s})$ describes
the data we define the quantile $\rm{RRM}_{X}$ via the relation
\begin{equation}
X=\int_{-\rm{RRM}_{X}\left(z\right)}^{+\rm{RRM}_{X}\left(z\right)}P(\rm{RRM},z)d\rm{RRM},
\end{equation}
that is the value such that a fraction $X$ of all the RRM's in a
distribution $P$ fulfills $\left|RRM\right|\leq \rm{RRM}_{X}$.
In Fig. $\ref{fig:quantiles}$ we plot both the observed data
(points) and the model (solid-line) as a function of redshift for the quantiles $\rm{RRM}_{0.68}$
(left) and $\rm{RRM}_{0.90}$ (right).

It is apparent that our simple model with magnetized strong MgII absorbers is
capable of reproducing the data distribution up to $z \sim 2$.  The
redshift dependence of the data in this model can be interpreted as
follows. At low redshifts very few lines of sight pass through an
intervenor so that $P(\rm{RRM},z_{s})$ is dominated by the intrinsic and
error components and varies only slowly with $z$.  As $z$ approaches $\sim 1$, however,
the probability of intersecting an absorbing system becomes significantly
higher. Here the role of intervenors starts to kick in and
RRM$_{0.68}$ grows gently.  As we move to sources at progressively higher
redshifts, the number of intervenors increases.  However, their
contribution is suppressed by the $(1+z)^{-2}$ dilution factor.  This
leads to a flattening in the distribution of RRM$_{0.68}$ at $z\sim
2$.  Similar reasoning applies to $\rm{RRM}_{0.90}$ except that the
few intervenors encountered at low redshifts are able to affect
this statistical quantity much earlier.

\begin{figure*}
\plotone{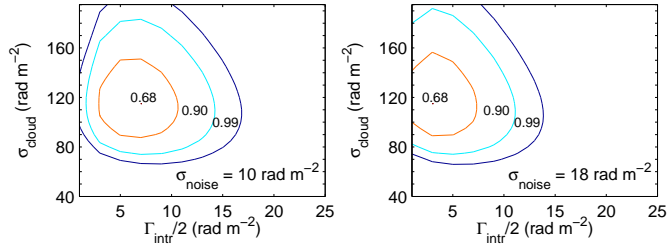}
\caption{Contours of constant likelihood as a function of fit
   parameters $\Gamma/2_{intr}$ and $\sigma_{cloud}$ for two fixed values of
   $\sigma_{noise}$. MgII systems with $W_{r} > 0.3\rm{\AA}$ are used as a model 
for the intervening systems.\label{fig:confidencelevel}}
\end{figure*}

To illustrate the robustness of our estimate of $\sigma_{cloud}$, we show in
Figure $\ref{fig:confidencelevel}$ 
contours of constant likelihood as a function of the
parameters $\sigma_{cloud}$ and $\Gamma_{intr}/2$ for two different
choices of $\sigma_{noise}$, which will bracket a good fraction of
the values this parameter can assume.  It is apparent that a change in
$\sigma_{noise}$ affects $\Gamma_{intr}/2$, and that these two quantities are
anticorrelated.  This makes sense in that $\Gamma_{intr}/2$ and
$\sigma_{noise}$ dominate at low redshift where, to some extent, their
role can be interchanged. Importantly, however, the uncertainties in
the error estimate do not seem to have an impact on $\sigma_{cloud}$,
which remains at about 3 $\sigma$ above the null value.

When we repeat the same analysis including the contribution of weak
absorbers, the quality of the model prediction, represented by the
dotted line in Fig. $\ref{fig:KStest}$, worsens, particularly for the quantity
$\rm{RRM}_{0.68}$.  
This is because the number of intervenors
encountered below redshift unity is substantially higher in the model, leading to a
higher expected value for $\rm{RRM}_{0.68}$.  This result may indicate
a limitation in our assumption of a column density independent
$\sigma_{cloud}$ for all absorbers. We expect that in attempting to reconcile
the data with the model, improving on this limitation would provide a better 
solution than postulating drastically different magnetic properties for weak and
strong MgII absorption systems. Also, establishing a $W_{\rm r}$ = 0
control sample, not available in this investigation, will improve
future parameter specifications for MgII interveners.  

Finally, we show a model in Figure $\ref{fig:quantiles}$ (dashed line) in which we 
set $\sigma_{cloud}=0$, i.e., which has only
an unchanging intrinsic component and a measurement error. 
The dashed lines in both panels of Fig.$\ref{fig:quantiles}$ show the quantiles
$\rm{RRM}_{0.68}$ and $\rm{RRM}_{0.90}$ predicted by this model, with
best fit parameters, in units of rad m$^{-2}$,
$\sigma_{noise}=13_{-3}^{+4}$ and $\Gamma_{intr}/2=21_{-6}^{+7}$. It clearly fails to
reproduce the behavior of the observed RRM statistics with redshift. The
reason this model fails so dramatically is that, since the noise component is
independent of redshift and the RM of the non-evolving intrinsic component 
declines as $(1+z)^{-2}$, it can only predict a decrease of both
quantiles as a function of redshift, instead of the observed increase.

\subsection{Implications for magnetic fields in early systems}
\label{sec:decline}

Whether due to intervenors or other evolution of galactic or pre-galactic
systems, the rise in the fraction of lines of sight affected by Faraday  
rotation at progressively higher redshifts indicates the presence of magnetized
gas in such systems at earlier epochs.
While the intervenor model would explain the behavior of the data with
the higher incidence of magnetized clouds, an alternative possibility
that we cannot yet exclude is that the RRM in the high
redshift sources is dominated by an intrinsic component that increases
into the past. Note that in this case, however, we require substantial {\it negative} evolution in the Faraday-active 
medium, i.e. with decreasing
$z$, in order not to over-predict the expected  
dispersion in the observed RRM values at low redshift (see Figure~\ref 
{fig:quantiles}).
In terms of our model parameters, for example, $\Gamma_{intr}$  
would have to decrease with time. This would follow naturally if the Faraday  
active medium were a magnetized cloud expanding as a result of being over$-$pressured  
with respect to the ambient medium.

In any case we can attempt to estimate the strength of the magnetic  
field in these earlier systems, given some independent estimate 
of the column density of free electrons, $N_{e}$.
Following Kronberg and Perry (1982), we can express an RRM$_C$ arising
at redshift $z_C$ as

\begin{equation}
\label{eq:RRMc}
{\rm RRM}_{C} = \beta (1 + z_{C} )^{ - 2} N_e \langle B_\parallel   
\rangle {\rm{
rad~m}}^{{\rm{ - 2}}},
\end{equation}
where $N_{e}$ is in cm$^{-2}$,
$\langle B_\parallel  \rangle$  accommodates any field reversal pattern, and 
is defined by
\begin{equation}
\langle B_\parallel  \rangle  = \frac{{\int\limits_{cloud} {n_e
B_\parallel  dl} }}{{\int\limits_{cloud} {n_e dl} }}.
\end{equation}
The constant $\beta$ = 2.63 $\times $10$^{-13}$ rad m$^{-2}$ cm$^ 
{2}$
Gauss$^{-1}$.

The rms deviation from zero of the observed RRM, $\sigma_ 
{RRM}$, can be straightforwardly
converted to an estimate of a typical $\langle B_\parallel \rangle$  
by inverting Eq.~($\ref{eq:RRMc}$) and inserting an estimate of $N_{e}$
for the system generating the Faraday rotation.
The important point is that as we move into the redshift range $1.8  
\la z \la 3.7$
the fraction of lines of sight, $f_{i}$, with near zero RRM, falls  
off sharply.
This is independent of whether the excess RRM is in the vicinity of the 
radio source (``intrinsic''), or located
in a galaxy system at intervening redshift.
Note that the exact redshift is not crucial for our estimate
here as the spread in  $(1 + z)^{2}$ over the interval $1.8\la z \la  
3.7$ is modest,
of order a few at most.

Inverting Eq.~(\ref{eq:RRMc}) and setting RRM$_C \sim \sigma_{\rm RRM}$
in the observer's frame (e.g. panel 9 of Fig.$\ref{fig:RRMhisto}$) gives
\begin{eqnarray} \nonumber
\label{eq:Bestim} \langle B_\parallel\rangle &=& 5.5\times10^{-7}  
{\rm G}
\left( \frac{1 +z_{C}}{3.5} \right)^{ 2} \\ & \times &
\left(\frac{\sigma_{\rm RRM}}{20\;{\rm rad~m}^{-2}}\right) \;
\left(\frac{N_e}{1.7\times 10^{21}{\rm cm}^{-2}}\right)^{-1}
\end{eqnarray}
for the system causing the RRM excess.

The local $<B_{||}$$>$ values within a system are expected to be larger because
of cancelations due to reversals
in the magnetic field orientation along the line of sight.
The column density assumed in the above normalization is
comparable to HII column densities through today's typical spiral  
galaxies. 
Thus if the RRM is due to galactic systems at high redshifts, our  
estimate
would imply the presence of magnetic fields there with strength of order
at least a $\mu$G, which is comparable to the value observed in the Galaxy.
If the radio source is embedded {\it within} a Faraday rotating cloud 
(the intrinsic RRM case), 
the $\sim $ 2 $\times$ lower RRM pathlength would, relative to a lower $z$ intervening cloud,
make the magnetic field strength correspondingly higher.
The results presented in \S 4.1 indicate that 
magnetic fields must have been generated very quickly in galaxy systems, at
early cosmological epochs.

\section{On the detectability of a widespread IGM magnetic field} \label{discussion:sec}

Limits on a widespread IGM magnetic field were first discussed and
derived in the 1970's under the assumption that the Universe was
homogeneous, had only baryonic matter, and was pervaded by contiguous
magnetized cells of comoving length, $l_{0}$. Under these assumptions, 
RM's measured at the largest redshifts (then ${\sim }$ 2)
placed upper limit estimates of $B^{\rm IGM}_{0}$ in the range
10$^{-8}$ - 10$^{-9}$ G (Rees \& Reinhardt (1972), Nelson (1973), Kronberg \&
Simard-Normandin (1976), Kronberg et al. (1977)). In this co-expanding
model in which $|B$($z$)$|$ $\propto$ (1 + $z$)$^{2}$, the observed
growth in RRM($z$) due to a widespread, co-expanding IGM is 
dominated by contributions at the largest redshifts. In the context of current 
concordance cosmology, and our knowledge of large scale 
filaments and voids of matter, these $B^{\rm IGM}_{0}$ limits are no 
longer appropriate, and need to be revisited.  

Now, isolating or limiting a widespread extragalactic RRM, and hence $B^{\rm IGM}_{0}$,
is more difficult. It requires that we disentangle a $B^{\rm IGM}$
from any cosmic evolution in RRM$_{C}$($z$) (case 2) and RRM$_{intr}$ 
(case 3). At least some
of the high redshift sources in our sample, such as 3C191 at $z$=1.95
(Kronberg et al. 1990), and several high-$z$ RM's measured by Carilli
et al.(1997), Athreya et al. (1998), and Pentericci et al. (2000)
indicate evolutionary effects in RRM$_{intr}$.

However the expected width of N(RRM$_{\rm{IGM}}$, $z$) due to a widespread, co-expanding
magnetized IGM increases steeply at the large redshifts (e.g. Kronberg, Reinhardt \& 
Simard-Normandin 1977). This raises the possibility that
future RRM datasets that extend to redshifts $z$ = 4 - 5 may 
be strongly influenced by a widespread co-expanding intergalactic medium. 

Then with better knowledge of the intervenor and source-intrinsic populations, 
RRM$_{C}$($z$) and RRM$_{intr}$($z$), it may be possible to explore, or limit 
the contribution of RRM$_{\rm IGM}$($z$). The data could then be compared with modelled 
contributions of RRM$_{\rm IGM}$($z$)(filaments) and 
RRM$_{\rm IGM}$($z$)(voids) based on LSS evolution simulations.   

A probe of the RRM contribution of some local universe galaxy filaments was 
recently attempted by Xu et al. (2006). Their initial estimate was 
$B_{\rm IGM} \sim$ 3 $\times$ 10$^{-7}$G in the 
Perseus-Pisces supercluster filament zones, scaled to an assumed field reversal scale
of $\sim$ 0.5 Mpc and to estimates of the electron density in the warm-hot IGM(WHIM).

\section{Summary and conclusions} \label{summary:sec}

We have analyzed an unprecedented large sample of RRM data 
extending to redshift 3.7.
The size of our new Faraday RM sample permits us to isolate
preferred Galactic ($l$,$b$) zones where Galactic foreground RM is
optimally removed with the help of 
a new, more accurate set of 1566 extragalactic source RM's that
are mostly off the Galactic plane. We find a clear 
increase with redshift of Galaxy-corrected RRM's at medium to low RRM levels, below
$\sim$ 100 rad m$^{-2}$ in the observer's frame. 
Our results and conclusions can be summarized as follows:

\begin{enumerate} 

\item{There is a striking and systematic decline,
significant at the $\sim$ 3 $\sigma$ level, in the fraction of sources
in the range $z\sim$ 1.5 - 2.3 that have $|$RRM$|$ less than 20 rad
m$^{-2}$. This is the first global indication that the Universe
becomes increasingly ``Faraday opaque'' to the highest redshift radio sources.}

\item{A model with intervening systems distributed according to the dN/dz statistics 
of strong MgII absorption line systems in QSO spectra is consistent with the observed growth in the widths 
of the RRM distribution up to $z\sim2$ with no evolution in the rest frame RM of each absorber. 
This hypothesis could be tested in the future by comparing the distribution of RM in quasar sightlines 
that have, and do not have, strong MgII absorption.}   

\item{We provide global estimates of early Universe magnetic field strengths in galaxy systems, 
based on the progressively increasing $``$Faraday opaqueness$"$ of systems beyond $z\sim 1.5$. 
These have at least $\mu$G level fields, and indicate that magnetic fields at these high redshifts 
were at least as strong as at the present epoch.}

\item{Redshift variations in the widths of RRM distributions below $z
\sim$ 1.2 are small and can only be specified at the 1.5 - 2$\sigma$
level, limited by residual uncertainties in the foreground Galactic RM. The
latter is at the level of $\sim 9 $ rad m$^{-2}$ in the higher $b$ zones selected, 
determined from a  new all-sky sample of 1566 RM's.}

\end{enumerate}

\section{Acknowledgements}

We thank Chris Carilli and Meri Stanley of NRAO for access to VLA data
that enabled us to calculate several additional high - $z$ RM's, and an anonymous referee
for helpful comments. PPK acknowledges support from the U.S. Department of Energy, the
Laboratory Directed Research and Development Program (LDRD) at Los
Alamos National Laboratory, and the Natural Sciences and Engineering
Research Council of Canada (NSERC). MLB is supported by the Swiss
National Science Foundation, and FM acknowledges support by the Swiss Federal
Institute of Technology through a Zwicky Prize Fellowship.


\begin{thebibliography}{}

\bibitem[Armengaud et al. (2005)]{armen05} Armengaud, E., Sigl, G., and Miniati, F. 2005 PRD {\bf 72}, 43009

\bibitem[Athreya et al. 1998]{athre98}
Athreya, R. M., Kapahi, V. K., McCarthy, P. J., \& van Breugel, W. 1998, \aap, 329, 809 - 820

\bibitem[Carilli et al. 1991]{cari91}
Carilli, C. L., Perley, R. A., Dreher, J. W., \& Leahy, J. P. 1991, ApJ,
383, 554

\bibitem[Carilli et al. 1997]{cari97}
Carilli, C. L., Roettgering, H. J. A., van Ojik, R., Miley, G. K., \& van
Breugel, W.J.M. 1997, ApJS, 109, 1

\bibitem[Churchill et al.(1999)]{church1999} Churchill, C.~W., Rigby, J.~R., Charlton, J.~C., \& Vogt, S.~S.\ 1999, \apjs, 120, 51

\bibitem[Dolag et al. (2005)]{dola05} Dolag, K., Grasso, D., Springel, V. and Tkachev, I. JCAP 01(2005) 009;

\bibitem[Kolatt 1998]{kol98}
Kolatt, T. 1998, \apj, 495, 564-579

\bibitem[Kronberg \& Perry 1982]{kro82}
Kronberg, P. P., \& Perry, J. J. 1982, \apj, 263, 518-532
L31-L34

\bibitem[Kronberg, et al. 1990]{kro90}
Kronberg, P. P., Perry, J. J., \& Zukowski, E. L. H. 1990, \apj, 355,
L31-L34

\bibitem[Kronberg, et al. 1992]{kro92}
Kronberg, P. P., Perry, J. J., \& Zukowski, E. L. H. 1992, \apj 387,
528-535

\bibitem[Kronberg \& Simard-Normandin 1976]{kro76}
Kronberg, P. P. \& Simard-Normandin, M. \ 1976 {\it Nature} 263, 653 - 656

\bibitem[Kronberg et al.(1977)]{kro77} Kronberg, P.~P., Reinhardt, M., \& Simard-Normandin, M.\ 1977, \aap, 61, 771

\bibitem[Kronberg et al. (2001)]{kron01} Kronberg, P.~P., Dufton, Q.~W., Li, H., \& Colgate, S.~A. 2001, \apj,560, 178

\bibitem[Mestel \& Paris (1984)]{mest84} Mestel, L., \& Paris, R. B. 1984, A\&A, 136, 98

\bibitem[Nelson (1973)]{nels73} Nelson, A.H. 1973, {\it Pub. Astron. Soc. Japan}, 25, 489

\bibitem[Nestor et al.(2005)]{nest2005} Nestor, D.~B., Turnshek, D.~A., \& Rao, S.~M.\ 2005, \apj, 628, 637

\bibitem[Oren \& Wolfe(1995)]{oren1995} Oren, A.~L., \& Wolfe, A.~M.\ 1995, \apj, 445, 624

\bibitem[Pentericci et al.(2000)]{penteric2000} Pentericci, L., Van Reeven, W., Carilli, C.~L., R{\"o}ttgering, H.~J.~A., 
\& Miley, G.~K.\ 2000, \aaps, 145, 121

\bibitem[Rees (1987)]{rees87} Rees, M. J. 1987, QJRS, 28, 197

\bibitem[Rees \& Reinhardt 1972]{rees72}
Rees, M. J. \& Reinhardt, M. 1972, \aap 19, 104,

\bibitem[Sigl et al. (2003)]{sigl03} Sigl, G., Miniati, F., and Ensslin, T. A. 2003 PRD 68, 043002

\bibitem[Sigl et al. (2004)]{sigl04} Sigl, G., Miniati, F., and Ensslin, T. A. 2004 PRD 70, 043007

\bibitem[Short et al (2007a)]{shorta07} Short, M. B., Higdon, D. M., \&
Kronberg, P. P. 2007a {\it Bayesian Analysis} (in press) {\it http://ba.stat.cmu.edu/forthcoming.php}

\bibitem[Short et al (2007b)]{shortb07} Short, M. B., Higdon, D. M., \&
Kronberg, P. P. 2007b {\it Bayesian Statistics} {\bf 8}, 665 - 660

\bibitem[Welter et al. 1984]{wel84}
Welter, G. L., Perry, J. J. \& Kronberg, P. P. 1984, \apj, 279, 19-39

\bibitem[Xu et al. 2006]{Xu06} Xu, Y., Kronberg, P. P., Habib, S., \& Dufton, Q. W. 2006 \apj, 637, 19

\bibitem[You et al. 2003]{You03} You, X. P., Han, J. L. \& Chen, Y. 2003 {\it Acta Astronomica Sinica}, 44, 155

\end{thebibliography}
\end{document}